# Quantum Evolution and Anticipation

by Hans-Rudolf Thomann

March 2010

Author address Im Altried 1h, 8051 Zürich, hr@thomannconsulting.ch

## CONTENTS

## Abstract

A state is said to exhibit orthogonal evolution of order $L$ and step size $T$ iff it evolves into orthogonal states at times $0 < |n|T \leq L$. Measurement at time $\frac{T}{2}$ yields states from this orthogonal set with certain probabilities. This effect, which we name anticipation, has been described in a previous paper in a special setting.

Here we extend our analyzes to general type quantum evolutions and spectral measures and prove that quantum evolutions possessing an embedded orthogonal evolution are characterized by positive joint spectral measure.

Furthermore, we categorize quantum evolution, assess anticipation strength and provide a framework of analytic tools and results, thus preparing for further investigation and experimental verification of anticipation in concrete physical situations such as the H-atom, which we have found to exhibit anticipation.

## 0 Introduction

This paper generalizes and extends the findings of (Thomann, 2008), where we studied the special case of orthogonal evolutions of maximal order with pure point and absolutely continuous measures, to arbitrary types of spectral measures and evolutions.

Under a Hamiltonian $H$, a quantum state $q$ evolves into an orbit $q_t = U_t q$, where $U_t = e^{-\frac{iH}{\hbar}t}$ is defined on the closure of $D(H)$. By spectral theory (Reed, 1980), the spectral measure $\mu_q(\lambda)$ of $q$ and $H$ is uniquely defined by $(q, f(H)q) = \int f(H) d\mu_q(\lambda)$ for any analytic function $f$.

For fixed *step size* $T > 0$, $q_0 = q$ and $n \in \mathbb{Z}$, an *evolution of order* $L \geq 0$ is given by the *primary sequence* $q_n = U_T^n q_0$ and the *dual sequence*, $r_n = U_T^n r_0$, such that $(r_k, q_l) = \delta_{kl}$ $(k, l \in \mathbb{Z}, |k - l| \leq L)$ (see definition 1). The amplitudes $(r_t, q_0)$ are determined by the spectral measure $\rho d\mu_q$ (see definition 2). *Positive evolutions* are defined as those with positive $\rho$.

Theorem 1 states that every positive evolution contains an *embedded orthogonal evolution*, driving the component $s_0$ of $q_0$ of size $\zeta$ through $L + 1$ mutually orthogonal states. Theorems 2 and 3 prove the existence of (not necessarily positive) evolutions of any order for point, singular-continuous and absolutely continuous spectrum, respectively.

*Quantum-Mechanical anticipation* (definition 3) is defined for the embedded orthogonal evolution in terms of the *anticipation amplitudes* $\alpha_n = (s_n, U_{T/2} s_0)$, and the *anticipation probabilities* $p_n = |\alpha_n|^2$, expressing the result of a measurement of $s_0$ at time $T/2$, where $|n| \leq L$. *Quantum-mechanical retrospection* is the time-reverse of anticipation. Measurements anticipate future (or recall past) states $s_n$ with probability $p_n$, which is the reason for the naming. Theorem 4 assesses the anticipation strength in various situations.

This paper provides analyses and tools facilitating further research about anticipation. In chapter 1 we introduce the basic terminology and definitions, characterize finite-dimensional and infinite-dimensional evolutions and their spectral measures, and explore the topology of the manifold of positive evolutions. Chapter 2 explains and defines anticipation, provides a representation of anticipation amplitudes and analyzes anticipation strength.

To cover not only point and absolutely continuous but also singular-continuous measures, we use some formalism from spectral theory an provide mathematical theorems in appendix A. Theorem A.1 establishes explicit formulae for the Fourier coefficients of continuous tions $F$ supported on $K$ whose second distributional derivatives yield a given measure $\nu$, inverting the discrete Fourier transform $d\nu^\wedge(n)$. Theorem A.2 proves a very general representation of Dirac's $\delta$ functional.

In a future paper we will present results from numerical simulations of the point spectrum of the H-atom as well as equidistant and random point spectra, demonstrating that anticipation and retrospection occur in common situations. The quantum anticipation simulator software running und Microsoft Windows is already downloadable from the author's homepage http://www.thomannconsulting.ch/public/aboutus/aboutus-en.htm.

# 1 Quantum Evolution Systematic

In this chapter we introduce the basic definitions and terminology, cast quantum evolutions into a set of categories and derive the characteristic spectral properties of each category.

## 1.1 Basic Terminology and Definitions

Let $H$ be an essentially self-adjoint Hamiltonian. The set of states we are considering is the Hilbert space $\mathcal{H}$ which equals the closure of $D(H)$. Throughout the paper all states will be assumed to have unit norm. New technical terms are introduced in *italic* type.

For any state $q \in \mathcal{H}$ the finite spectral measure $\mu_q(\lambda)$ is uniquely defined by the inner product $(q, f(H)q) = \int f(H) d\mu_q(\lambda)$ for any analytic function $f$ (see e.g. (Reed, 1980)) This relation induces a unique unitary mapping $\mathfrak{J}_q: \mathcal{H} \to \mathcal{L}^2(d\mu_q)$ such that the image of $q$ is the constant function $1$. We will sometimes identify states with their images under this mapping.

The evolution operator $U_t = e^{-\frac{iH}{\hbar}t}$ is a unitary isomorphism of $\mathcal{H}$, under which $q$ evolves into an orbit $q_t = U_t q$. The amplitude $(q_0, q_T)$ equals the Fourier transform $d\hat{\mu}_q(T) = \int e^{-i\lambda T/\hbar} d\mu_q(\lambda)$. By the unitarity of $U_t$, $(q_0, q_T) = (q_t, q_{t+T})$ is independent from $t$.

Our subject is the evolution of states at equidistant time steps. To analyze them we define evolution scenarios comprising a primary and an associated dual sequence of states.

Definition 1: Evolution Scenario

An *evolution scenario* (or short: evolution) is characterized by the following elements:

a) A Hamiltonian $H$.
b) The *step time* $T > 0$.
c) The *primary base state* $q_0$.
d) The doubly-infinite *primary sequence* $q_n = U_T^n q_0$ $(n \in \mathbb{Z})$.
e) The *primary amplitudes* $\beta_n = (q_0, q_n)$ satisfying $\beta_{-n} = \overline{\beta_n}$.
f) The cyclic space $C = \mathrm{Span}(\{q_n | n \in \mathbb{Z}\})$, which is invariant under $U_T$.
g) The *dimension* $d = \dim(C)$, $0 < d \le \infty$.
h) The integer *order* $L$, $d > L \ge 0$.
i) The subspace $C_L = \mathrm{Span}(\{q_n | 0 \le |n| \le L\})$ of $C$.
j) The *dual base state* $r_0 \in C_L$ of unit norm meeting $(r_0, q_n) = \delta_n$ $(n \in \mathbb{Z}, 0 \le |n| \le L)$ and $(r_k, q_l) = \overline{(r_l, q_k)}$ $(k, l \in \mathbb{Z})$.
k) The doubly-infinite *dual sequence* $r_n = U_T^n r_0$ $(n \in \mathbb{Z})$ in $C$.

An evolution scenario is said to exhibit *orthogonal evolution of order* $L$ iff $q_n = r_n$.

■

Evolution scenarios exist for any $U_T$, $q_0$ and $0 < L \le \lfloor (d-1)/2 \rfloor < \infty$, as we will prove in theorems 2 and 3. For $L = \infty$ see theorem 2e).

Provided its existence $r_0$ is that vector in the orthogonal complement of the hyperplane $\mathrm{Span}(\{q_n | 0 < |n| \le L\})$ relative to $C_L$ uniquely defined by the requirement that $(r_0, q_0) = 1$, and $\infty > \|r_0\|_2 \ge 1$.

As, by the unitarity of $U_t$, $(r_0, q_n) = (r_k, q_{k+n})$ is independent from $k$, our construction ensures that $(r_{n+k}, q_k) = \delta_n$ $(n, k \in \mathbb{Z}, 0 \leq |n| \leq L)$. Thus, within the order, the dual sequence is orthogonal to the primary sequence.

In case of orthogonal evolution, the primary and dual sequence coincide, thus $(r_0, q_0) = \|q_0\|_2^2 = 1$.

State evolution is completely determined by spectral measures and quantities, which we introduce now.

Definition 2: Spectral measures and quantities

Until the end of this chapter $\|\cdot\|_n$ denotes the $\mathcal{L}^n(d\mu_q)$ norm. The Lebesgue norm is $\|\cdot\|_{n,d\lambda}$.

a) The spectral measure of the primary and dual sequence are denoted by $\mu_q(\lambda)$ and $\mu_r(\lambda)$, respectively.
b) With $\rho = \mathfrak{J}_q(r_0) \in \mathcal{L}^2(d\mu_q)$ the *joint spectral measure* $\mu_s(\lambda)$ is $d\mu_s = \frac{|\rho|}{\|\rho\|_1} d\mu_q$.
c) *Positive evolution* is characterized by $\rho$ being a non-negative function.
d) $\mathrm{Shift}(\kappa) = \begin{cases} \kappa & (0 \leq \kappa < \pi) \\ \kappa - 2\pi & (\pi \leq \kappa < 2\pi) \end{cases}$ maps the interval $[0,2\pi)$ to $K = [-\pi,\pi)$.
e) For any finite measure $\mu_x$ and real $m$, the *reduction modulo* $m$ is $d\mu_{x,m}(\kappa) \stackrel{\text{def}}{=} \sum_{\kappa=\lambda \bmod m} d\mu_x(\lambda)$.
f) For $x = q, r, s$, the reduced measure $\nu_x$ is defined by $d\nu_x(\mathrm{Shift}(\kappa)) = d\mu_{x,2\pi}(\kappa)$.
g) $1' = \mathrm{sign}\,\rho$.
h) $\mathfrak{J}_q(s_n) = \sqrt{\frac{|\rho|}{\|\rho\|_1}} e^{-i\lambda nT/\hbar}$, $\mathfrak{J}_q(s_n') = 1'\sqrt{\|\rho\|_1|\rho|}$.
i) $\zeta = (s_0, q_0)$.

■

$\mu_x(\mathbb{R}) = \|x_0\|_{2,d\lambda}^2 = 1$ for $x = q, s$, and the supports of these measures are equal up to a set of zero measure.

Obviously $d\mu_r = |\rho|^2 d\mu_q$, $\mathfrak{J}_q(q_n) = e^{-i\lambda nT/\hbar}$, $\mathfrak{J}_q(r_n) = \rho e^{-i\lambda nT/\hbar}$ and $(q_0, r_n) = \int \rho e^{-i\lambda nT/\hbar} d\mu_q = (\rho d\mu_q)^{\wedge}(nT/\hbar)$.

The definition of $\mu_s$ requires $\|\rho\|_{1,d\mu_q} < \infty$, which indeed follows from the Schwartz inequality and $\|1\|_2 = \mu_q(\mathbb{R}) = 1$: $\|\rho\|_1 = \|\rho \cdot 1\|_1 \leq \|\rho\|_2\|1\|_2 = \|\rho\|_2 = \|r_0\|_{2,d\lambda} < \infty$.

In case of positive evolution, $\|\rho\|_1 = \int \rho\, d\mu_q = (r_0, q_0) = 1$, $d\mu_s = \rho d\mu_q$ and $\mathfrak{J}_q(s_0) = \rho^{1/2}$, otherwise $\|\rho\|_1 > 1$.
$\zeta = \left\|\sqrt{\frac{|\rho|}{\|\rho\|_1}}\right\|_1 = \|s_0\|_1$ is a positive quantity. $0 < \zeta \leq 1$, as it is the inner product of two unit vectors. In case of positive evolution $\zeta = \int \rho^{1/2}\, d\mu_q$.

The definition of $1'$ requires that $\rho$ be real-valued. This holds because $(q_0, r_n + r_{-n}) = 2\rho \cos \lambda nT/\hbar$, $(q_0, r_{-n} - r_n) = 2\int \rho \sin \lambda nT/\hbar\, d\mu_q$ and $(r_n \pm r_{-n}, q_0) = 0$ $(0 < |n| < L)$. As $\int \rho\, d\mu_q = 1 > 0$, $\rho$ is the unique solution of a system of purely real (integral) equations.

$d\mu_s = \frac{|\rho|}{\|\rho\|_1} d\mu_q$ is a positive measure, $(q_0, r_n) = \|\rho\|_1 \int 1' \, e^{-i\lambda nT/\hbar} d\mu_s = (s_0, s_n')$. In case of positive evolution $1' = 1$ a.e., $\|\rho\|_1 = 1$, $s_n' = s_n$ and $s_n$ has orthogonal evolution.

$\rho$ is periodic with period $2\pi \hbar/T$, as it has a representation $\rho = \sum_{|n|\leq L} c_n e^{-i\lambda nT/\hbar}$ for certain coefficients $c_n$, which satisfy $c_n = \overline{c_{-n}}$ because $\rho$ is real-valued. The same follows for $1'$ and $\mathfrak{I}_q(s_n)$.

The unitary mappings $\mathfrak{I}_x: \mathcal{H} \to \mathcal{L}^2(d\mu_x)$, existing for $x = q, s$, induce unitary mappings $\varkappa_x: C \to \mathcal{L}^2(d\nu_x)$ between the cyclic space $C$ and the Lebesgue spaces of the reduced spectral measures. Due to the periodicity of $\rho$, $1'$ and $\mathfrak{I}_q(s_0)$ there holds $\varkappa_q(r_n) = \rho e^{-i\lambda nT/\hbar}$ in $\mathcal{L}^2(d\nu_q)$, $d\nu_s = \frac{|\rho|}{\|\rho\|_1} d\nu_q$ and $\varkappa_s(s_n) = e^{-i\lambda nT/\hbar}$ in $\mathcal{L}^2(d\nu_s)$, furthermore $\varkappa_s(q_0) = \sqrt{\frac{\|\rho\|_{1,d\nu_q}}{|\rho|}}$ and $\varkappa_s(r_0) = 1'\sqrt{\|\rho\|_1 |\rho|}$, as well as $\|\rho\|_n = \|\rho\|_{n,d\nu_q}$.

The reduced spectral measures as well determine the state evolution at integer times: $(q_0, q_n) = d\nu_q^\wedge(nT/\hbar)$ and $(r_0, q_n) = \|\rho\|_1 d\nu_s^\wedge(nT/\hbar)$. The proof is given in the following proposition which summarizes the above findings.

Until stated otherwise, we assume below $T/\hbar = 1$ and omit this factor from all formulae.

Proposition 1: Spectral Measures

a) The measures $\mu_x$ $(x = q, s)$ are positive and have total measure $\mu_x(\mathbb{R}) = 1$. $d\mu_r = |\rho|^2 d\mu_q$, $\mathfrak{I}_s(q_0) = |\rho|^{-1/2}\|\rho\|_1^{1/2}$ and $\mathfrak{I}_s(r_0) = 1'|\rho|^{1/2}\|\rho\|_1^{-1/2}$.
b) $\rho$, $1'$ and $\mathfrak{I}_q(s_0) = |\rho|^{1/2}\|\rho\|_1^{-1/2}$ are real, periodic functions.
c) The measures $\nu_x$ $(x = q, s)$ are positive and have total measure $\nu_x(K) = 1$. $d\nu_s = \|\rho\|_1^{-1}|\rho| d\nu_q$, $\varkappa_s(q_0) = |\rho|^{-1/2}\|\rho\|_1^{1/2}$ and $\varkappa_s(r_0) = 1'|\rho|^{1/2}\|\rho\|_1^{-1/2}$.
d) The state evolution at integer times is determined by the reduced measures: $(q_0, q_n) = d\nu_q^\wedge(n)$ and $(s_0, s_n') = \|\rho\|_1 (1' d\nu_s)^\wedge(n)$.
e) In case of positive evolution $s_n$ has orthogonal evolution, $\|\rho\|_1 = 1$, $\mathfrak{I}_q(s_0) = \mathfrak{I}_q(s_0') = \rho^{1/2}$ and $\zeta = \int \rho^{1/2} \, d\mu_q$.
f) $q_n$ has orthogonal evolution of order $L \geq 0$ iff $\zeta = 1$ iff $\nu_q = \nu_r = \nu_s$.

Proof:
For parts a-c) and e-g) see the above derivations. Part d) follows from

$$(q_0, q_n) = \int e^{-i\lambda n} \, d\mu_q(\lambda) = \sum_{m\in\mathbb{Z}} \int_{\lambda=2\pi m}^{2\pi(m+1)} e^{-i\lambda n} \, d\mu_q(\lambda) = \int_{\kappa=-\pi}^{\pi} e^{-i\kappa n} \sum_{\kappa=\text{Shift }(\lambda \bmod 2\pi)} d\mu_q(\lambda)$$
$$= d\nu_q^\wedge(n)$$

■

From this proposition there follows a simple, but very important observation, which we state in our first theorem.

Theorem 1: Embedded Orthogonal Evolution

Let $q_0$ have positive evolution of order $L$. Then there is a component of $q_0$ which has orthogonal evolution of order $L$. This component is $\zeta s_0$ and has norm $0 < \zeta \leq 1$. Furthermore, $\operatorname{Supp} \mu_s \subset \operatorname{Supp} \mu_q$.

The evolution of $s_0$ is the *embedded orthogonal evolution of* $q_0$, and $\zeta$ is its *size*.

Proof: By proposition 1, $s_0$ has positive evolution. The inclusion of the supports is proper, as e.g. a spectral measure of dimension $d$ whose support contains an equidistant point spectrum of dimension $L + 1 \leq d$ has an embedded orthogonal evolution of order $L$.

■

Until the end of this chapter we will, based on proposition 1d, confine to the reduced spectral measures.

## 1.2 Characterization of Finite-dimensional Evolution

In the finite-dimensional case the cyclic space and the Lebesgue spaces have finite dimension. The spectrum is pure point, and the reduced spectral measure has support on $d$ points. This occurs if either the underlying state space $\mathcal{H}$ is finite-dimensional, or if the support of $\mu_q$ has a regular structure such that the reduction yields only finitely many points. An example is the reduction of the spectrum of the Schrödinger operator $-\Delta + X^2$ modulo a rational number.

### 1.2.1 Amplitudes, Characteristic Polynomial, Eigenvalues and Spectra

In this section we explore the relationships of amplitudes, characteristic polynomial, eigenvalues and spectra.

Let $\kappa_n$ $(0 \leq n < d)$ be the ordered sequence of the reduced spectrum, $\omega_n = e^{-i\kappa_n}$ the eigenvalues of $U_T|C$, and $w_n = \nu_q(\{\kappa_n\})$ the *weights*.

Finiteness of the dimension implies $q_d = \sum_{0 \leq n < d} a_{d-n} q_n$, or equivalently $U_T^d q_0 = \sum_{0 \leq n < d} a_{d-n} U_T^n q_0$. As $q_0$ can be replaced by any other base vector of $C$, we conclude $U_T^d|C = \sum_{0 \leq n < d} a_{d-n} U_T^n|C$. The eigenvalues of $U_T|C$ therefore are roots of the characteristic polynomial

(1) $$\omega_k^d = \sum_{0 \leq n < d} a_{d-n} \omega_k^n \ (0 < k < d).$$

Notice that the coefficients $a_k$ have only $d$ real degrees of freedom, as they are functions of the $\omega_k$ and thus the $\kappa_n$. Indeed, $(-1)^n a_n$ is the basic symmetric function of order $n$ of the $\omega_k$, particularly $|a_d| = 1$, $a_n = a_d \bar{a}_{d-n}$ and thus $|a_n| = |a_{d-n}|$.

The weights $w_n$ are determined by the primary amplitudes from the equation system

(2) $$\beta_k = \sum_{0 \leq n < d} w_n \omega_n^k \ (0 \leq k < d).$$

Given $2d - 1$ primary amplitudes, the coefficients of the characteristic polynomial are easily found from the linear system

(3) $\beta_{d+k} = \sum_{0\leq n<d} a_n\beta_{n+k}$ $(0 \leq k < d)$,

However, $\beta_0, \dots, \beta_{\lfloor d/2 \rfloor}$ contain $d$ real degrees of freedom and are sufficient to determine $\nu_q$, as the $a_k$ are as well solutions of the non-linear system obtained from (3) by the replacement $a_n = a_d\bar{a}_{d-n}$ $\left(0 < n \leq \left\lfloor\frac{d}{2}\right\rfloor\right)$ and $\beta_{-n} = \bar{\beta}_n$.

Once the characteristic polynomial is found, (1) yields the eigenvalues and (2) the weights.

For any $0 < L \leq \lfloor(d-1)/2\rfloor$ and any $\nu_q$ there are $d-1-2L$ linearly independent real solutions $\rho_n = \rho(\kappa_n)$ satisfying the system

(4.1) $\delta_k = \sum_{0\leq n<d} \rho_n w_n \cos k\kappa_n$ $(0 \leq k \leq L)$

(4.2) $0 = \sum_{0\leq n<d} \rho_n w_n \sin k\kappa_n$ $(0 < k \leq L)$

Now $\nu_r$ and $\nu_s$ are obtained from proposition 1c). $\|\rho\|_1(1'd\nu_s)^\wedge(n) = \delta_n$ $(|n| \leq L)$ as required by definition 1j). Thus $\nu_q$ and $L \leq \lfloor(d-1)/2\rfloor$ determine the $d-1-2L$-dimensional linear space $D_{d,L}$ of solutions of (4.1-2) and the corresponding evolution scenarios, where $\nu_r$ and $\nu_s$ obey (4.1-2) and satisfy definition 1j).

$\nu_s$ and $\zeta$ do not generally completely determine $\nu_q$, as we will show in theorem 2.

For $L > \lfloor(d-1)/2\rfloor$ the system (4) is over-determined. (1) and (4.1) imply $(r_0, q_{L+1}) = (r_0, \sum a_k q_{L+1-k}) = a_{L+1}$ and $(r_0, q_{-L-1}) = (r_{L+1}, q_0) = (\sum a_k r_{L+1-k}, q_0) = \overline{a_{L+1}}$. If $a_{L+1} = 0$, then definition 1j) is met and the evolution scenario has order $L+1$. Induction and $|a_n| = |a_{d-n}|$ imply $a_L = \dots = a_{d-L} = 0$. Thus, for special sets of eigenvalues, (4) has a solution, and evolution scenarios with $\lfloor(d-1)/2\rfloor < L < d$ do exist.

In case of essentially periodic motion, $U^p = e^{i\vartheta}$ for some $\vartheta$, and the $\omega_n$ are essentially[1] $p^{\text{th}}$ unit roots. If $p = d < \infty$, then the eigenvalues form a complete set of unit roots, and $L = d-1$ iff $d\nu_s$ has equal weights $\sigma_n \stackrel{\text{def}}{=} \nu_s(\{\kappa_n\}) = p^{-1}$, as this is the unique solution of (4). In turn, evolution of $L = d-1$ is only possible in the minimal periodic case $p = d$, because equation (4) implies $\sigma_n = p^{-1}$.

### 1.2.2 Domain of Positivity

Let $d < \infty$ and $\sigma_n \stackrel{\text{def}}{=} \nu_s(\{\kappa_n\})$. The *domain of positivity of dimension $d$ and order $L$* is the set of eigenvalues $\kappa_n$ admitting positive $d$-dimensional evolutions of order $L$, $P_{d,L} = \{0 \leq \kappa_0 < \dots < \kappa_d \neq \kappa_0 < 2\pi | \sigma_n > 0\ (0 \leq n < d)\}$, where the $\sigma_n$'s satisfy

(4.4) $\delta_k = \sum_{0\leq n<d} \sigma_n \omega_n^k$ $(0 \leq |k| \leq L)$.

$P_{d,0}$ is the convex sub-set of the $d$-cube of edge length $2\pi$ defined by the constraint $0 \leq \kappa_0 < \dots < \kappa_d \neq \kappa_0 < 2\pi$. $P_{d,L} \subset P_{d,L-1}$ by definition 1j).

If $0 \leq L \leq \lfloor(d-1)/2\rfloor$, then $P_{d,L}$ is an open manifold of dimension $d$, because the system (4) is always solvable and establishes in the interior of $P_{d,L}$ a continuous (even analytic) relationship between the $\sigma_n$'s and the $\kappa_n$'s. Due to continuity, in some neighborhood of any element of $P_{d,L}$ the $\sigma_n$'s remain positive.

[1] i.e. up to a phase shift by $\vartheta/p$

If $d > L > \lfloor (d-1)/2 \rfloor$, then $P_{d,L}$ is an open manifold of dimension $2(d-L)-1$, because in this case, as we have seen in the preceding section, some $a_n = 0$, and each of these complex equations constrains two real degrees of freedom.

The boundary of $P_{d,L}$ consists on the one hand of the manifold $M_{d,L}$ of eigenvalues admitting at least one non-negative but no positive solution of order $L$, and on the other hand of degenerate elements with one or more degenerate pairs $\kappa_n = \kappa_m$. By continuity, the $\sigma_n$'s either converge to positive limits as elements of $P_{d,L}$ are degenerating, or the degenerate elements are lying on $M_{d,L}$.

In the former case we identify the degenerate element with an element of $P_{e,L}$, where $L < e < d$, as $P_{e,L} = \emptyset$ for $e \leq L$. Each element of $P_{e,L}$ corresponds to $e(d-e)$ degenerate elements of $P_{d,L}$.

Limit points of $P_{d,L}$ (degenerate and non-degenerate) with some zero $\sigma_n$'s are lying on $M_{d,L}$. Each element of $M_{d,L}$ can be constructed as an element of $P_{e,L}$ combined with $d-e$ points in $[0,2\pi]$ of zero weight, where $L < e < d$.

Let $N_{d,L} = P_{d,0}$ for $L \leq \lfloor (d-1)/2 \rfloor$, and the sub-set of $P_{d,0}$ satisfying the special constraints mentioned above in case of $L > \lfloor (d-1)/2 \rfloor$. Then $P_{d,L}$ is open in the topology of $N_{d,L}$, and $P_{d,L} \cup M_{d,L} \subset N_{d,L}$. $P_{d,L} \subset P_{d,L-1}$, as $P_{d,L}$ satisfies all constraints that $P_{d,L-1}$ does. $M_{d,L}$ partitions $P_{d,L-1} \cap N_{d,L}$ into $P_{d,L}$ and $\left(P_{d,L-1} \cap N_{d,L}\right) - P_{d,L}$.

### 1.2.3 Summary

The following theorem summarizes in parts a-c and e-f the results of the preceding two sections and completes them in parts d and g. It does however not reproduce the full details and all finesses of the above discussion.

Theorem 2: Characterization of Finite-dimensional Evolution

If $d < \infty$, then the reduced spectrum is pure point and

a) $\nu_q$ is completely determined by $\beta_k$ $(0 \leq k \leq \lfloor d/2 \rfloor)$.
b) Evolution scenarios exist for any $\nu_q$ and $0 < L \leq \lfloor \lfloor (d-1)/2 \rfloor \rfloor$, as well as for $\lfloor (d-1)/2 \rfloor < L \leq d$ provided that $a_n = 0$ $(\lfloor (d-1)/2 \rfloor < n < L)$.
c) The solutions of the system (4.1-2) form the $d-1-2L$-dimensional linear space $D_{d,L}$. Each solution completely determines $\nu_r$, $\nu_s$ and $\zeta$.
d) If $d > 2$, then for any $\nu_s$ and $0 < \zeta < 1$ there are several solutions $\nu_q$, $\nu_r$.
e) $L = d-1$ iff the motion is essentially periodic with $p = d$ iff $\sigma_n = p^{-1}$.
f) The domain of positivity, $P_{d,L}$, is an open manifold of dimension $d$ $(d > 2L \geq 0)$ and $2(d-L)-1$ $(2L \geq d > L > 0)$, respectively, nested into $P_{d,L-1}$. Its boundary is constructible from $P_{e,L}$, where $L < e < d$.
g) Necessary and sufficient for any measure $\nu_q$ to admit non-negative solutions of order $L = 1$ is that $\sup \operatorname{Supp} \nu_q - \inf \operatorname{Supp} \nu_q > \pi$.

Proof:
For parts a-c and e-f) see the derivations above.
Part d) We prove the proposition for $1 < d \leq \infty$ and arbitrary spectrum. Let the support of the reduced measures be partitioned into two sets $A, B$ with non-zero measure $a, b$ with respect to $d\nu_s$, respectively. Then there is exactly one solution such that $\mathfrak{I}_s(q_0)$ has constant value $0 < x \leq 1,\ y \geq 1$ on $A, B$, respectively, because proposition 1 implies $ax + by = \zeta$ and $ax^2 + by^2 = 1$, which yields a quadratic equation with non-negative discriminant, such that there is a unique solution meeting the inequalities. The claim follows from the fact that in the case of pure point spectrum there are $2^d - 2$ partitions satisfying the premises.
Part g) One verifies this criterion easily in case of point spectrum by noticing that it guarantees opposite sign of the minimum and maximum terms. The extension to continuous spectrum is straight-forward.

■

### 1.3 Characterization of Infinite-dimensional Evolution

In the infinite-dimensional case the cyclic space and the Lebesgue spaces have infinite dimension. The spectrum is a mixture of absolutely continuous, singular-continuous and discontinuous spectrum. Infinite-dimensional reduced point spectrum has isolated or even dense points of accumulation. An example for the latter is the reduction of the spectrum of $X^2$ modulo an irrational number. However, the formulae of theorem 2 below hold even in this case.

Relationships similar to those in the finite-dimensional case hold between the various quantities, with slight differences.

Theorem 3: Characterization of Infinite-dimensional Evolution

If $d = \infty$, the reduced spectrum is a mixture of absolutely continuous, singular-continuous and discontinuous spectrum. The motion is aperiodic.

a) $\nu_q$ is completely determined by $\beta_n\ (n \in \mathbb{Z}, n \geq 0)$, $d\nu_q = D^2F - \beta_0\delta_\pi - \gamma_0 D\delta_\pi$ for the continuous function $F$ and constant $\gamma_0$ given by Lemma A.1 below.
b) If $\nu_q$ has dense support in $I = [\kappa_0, \kappa_\infty]$, then for any $0 < L < \infty$ there is a solution $\rho$.
c) For any real measure $\nu_s$, $0 < \zeta < 1$ and $L < \infty$ there are infinitely many solutions $\nu_q$.
d) $\zeta = 1$ implies orthogonal evolution $\nu_q = \nu_r = \nu_s$ of length $L \leq \infty$.
e) $L = \infty$ implies absolutely continuous spectrum $d\nu_s(\kappa) = \frac{1}{2\pi} d\kappa$ on $K$, while $0 < \zeta \leq 1$.
f) Let $\nu_q$ have dense support in $\cup[\kappa_n - \varepsilon, \kappa_n + \varepsilon]$, where the $d$-tuple of $\kappa_n$'s is an element of $P_{d,L}$. Then positive evolution occurs for sufficiently small $0 < \varepsilon \ll L^{-1}$.

Proof:
Part a) From $\beta_{-n} = \bar{\beta}_n$ and Lemma A.1 below.

Part b) Partition the support of $\nu_q$ into $2L + 1$ intervals $I_m$, define $c_{nm} = \int_{I_m} e^{-in\kappa}\, d\nu_q(\kappa)$ and solve $\sum_m \rho_m c_{nm} = \delta_n$, $\sum|\rho_m|^2 = 1$. Then $\rho(\kappa) = \rho_m\ (\kappa \in I_m)$ is a solution.

Part c) The proof of theorem 1b) applies and yields the claim.
Part d) By proposition 1f.
Part e) By Lemma A.1, the given solution (see example A.1) is unique.

Part f) Partition the support into $N$ intervals and determine $\rho$ as in part b). As $N \to \infty$, the solutions approximate those of the finite-dimensional problem, and the openness of $P_{d,L}$ guarantees positivity.

■

# 2 Anticipation in Quantum Evolution

In this chapter we formally define anticipation for positive evolutions, establish a representation of the anticipation amplitudes in terms of the spectral difference and set up the statistical model for the assessment of anticipation in the remaining chapters.

## 2.1 Anticipation and Retrospection

Definition 3: Anticipation Amplitudes and Probabilities

Consider a positive evolution of order $L$. Then $s_0$ has orthogonal evolution of order $L$, and for $0 \leq n \leq L$ and $0 < \zeta \leq 1$

a) $\alpha_n = \zeta\left(U_{nT}\, s_0, U_{-T/2} s_0\right) = \zeta d\mu_s\left(n + \frac{1}{2}\right) = \zeta \int_{-\pi}^{\pi} e^{-i\left(n+\frac{1}{2}\right)\kappa} d\mu_s(\kappa)$ is the $n^{\text{th}}$ *anticipation amplitude*

b) $p_n = |\alpha_n|^2$ is the $n^{\text{th}}$ *anticipation probability*.

■

Let $A$ be an observable defined on $\mathrm{Span}\{s_n | 0 \leq |n| \leq L\}$ (whose dimension lies in the range $[L+1, d]$), such that $s_n$ $(0 \leq n \leq L)$ is an eigenfunction of $A$. Then $p_n$ is the probability to obtain $s_n$, when measuring $A$ at time $-\frac{T}{2}$. Thus $p_n$ is the probability to observe at time $-\frac{T}{2}$ that component of the embedded orthogonal evolution appearing in the sequence of discrete evolution steps not before the $n^{\text{th}}$ step: With probability $p_n$, measurement at time $-\frac{T}{2}$ anticipates the state at time $nT$, and projects the target into state $s_n$.

This is the reason to name the effect anticipation. Due to anticipation, measurement interactions can speed-up processes and de-stabilize states. Let's define "computation" as "controlled state preparation by quantum evolution". Then due to anticipation measurements at time $-\frac{T}{2}$ may provide a rapid advance of computation results.

If $B$ is an observable with eigenfunctions $s_n$ $(-L \leq n \leq 0)$, then measurement of $B$ at time $+\frac{T}{2}$ will project the state back to a previous state. This effect, which we name retrospection, can slow-down processes and stabilize states. Retrospection amplitudes and probabilities can be defined analogous to definition 3 above, and all of the following analyzes of anticipation hold as well for retrospection.

In case of non-positive evolution measurements are less likely to project states forward or backwards on their trajectory, as there is no embedded orthogonal evolution. Therefore we confine our definition and analysis of anticipation and retrospection to positive evolutions. The occurrence of positive evolutions will be investigated in a future paper.

## 2.2 Anticipation Amplitudes and the Spectral Difference

In chapter 1 we have seen that the evolution is determined by the reduced spectra. The anticipation amplitudes being defined at half-step time thus can be represented by the reduced spectra of the evolution with step size $1/2$. In definition a) below the modular factor selects from the real axis the odd and even numbered intervals of length $2\pi$, respectively.

Recall from (Last, 1996) that the measure $g$ is absolutely continuous with respect to the measure $h$, iff $dg = f dh$, for some Borel function $f$.

Definition 4: Spectral Difference

a) $dx_{s,j}(\text{Shift}(\kappa)) = d\mu_{s,4\pi}(\kappa + 2\pi j)$ $(j = 0,1)$ $(0 \leq \kappa \leq 2\pi, j = 0,1)$.
b) $dx_s = dx_{s,0} - dx_{s,1}$ is the *spectral difference measure*.
c) $\mu$ is *well-behaved* iff $dx_{s,i} = g_i dh\ a.e.$ for some measure $h$ and Borel functions $g_i$.
d) Let $\mu$ be well-behaved. Then $y = \frac{g_0 - g_1}{g_0 + g_1}$ is the *spectral difference function*.

See definition 2d-e) for the notation used in a).

■

In the sequel, we require $\mu$ to be well-behaved, where $h$ is admitted to be a spectral measure of arbitrarily general type. The following lemma shows that this is a weak condition.

Lemma 1: Well-behaved measures

a) If $\mu$ is pure point, then it is well-behaved.
b) If $\mu$ is absolutely continuous, then it is well-behaved.
c) If the support of $dx_s$ can be partitioned into measurable sets, such that on each set the restrictions of $dx_{s,0}$ and $dx_{s,1}$ to this set are normally related, then $\mu$ is well-behaved. Two measures, $f$ and $g$, supported on a set $S$, are said to be normally related if either $f$ is absolutely continuous w.r.t. $g$, or vice versa, or $S$ has zero measure under one of them.

Proof:

a) $dx_{s,0}$ and $dx_{s,1}$ are pure point and meet the premises of c).
b) $dx_{s,0}$ and $dx_{s,1}$ are absolutely continuous and meet the premises of c).
c) Consider a set $S$ and the restriction of the measures to this set. If $dx_{s,0}$ is absolutely continuous w.r.t. $dx_{s,1}$, then set $dh = dx_{s,1}$, $g_1 = 1$, $g_0 = \frac{dx_{s,0}}{dx_{s,1}}$. If $dx_{s,0}$ is singular w.r.t. $dx_{s,1}$, then set $dh = dx_{s,0}$, $g_0 = 1$, $g_1 = 0$. The reversed cases are analogous.

■

In the well-behaved case the following representations hold:

Proposition 2: Representation of Anticipation Amplitudes

a) $dx_{s,0} + dx_{s,1} = dv_s$, $dx_s = y dv_s$.
b) $\alpha_n = dx_s^{\wedge}\left(n + \frac{1}{2}\right) = \int_{-\pi}^{\pi} e^{-i\left(n+\frac{1}{2}\right)\kappa} dx_s(\kappa)$.
c) $y$ is a real function in $\mathcal{L}^2(dv_s)$, $|y| \leq 1$.
d) $\alpha_n = \zeta (y dv_s)^{\wedge}\left(n + \frac{1}{2}\right)$.

Proof:

a) By definition 2 and 4) and the well-behavior of $\mu$.
b)

$$\alpha_n = \zeta \int_{-\pi}^{\pi} e^{-i\left(n+\frac{1}{2}\right)\kappa} dx_{s,0} + e^{-i\left(n+\frac{1}{2}\right)(2\pi+\kappa)} dx_{s,1}$$

$$= \zeta \int_{-\pi}^{\pi} e^{-i\left(n+\frac{1}{2}\right)\kappa}\left(dx_{s,0} - dx_{s,1}\right).$$

c) By definition 4d).
d) Straight-forward from part a-b) and definition 2.

■

### 2.3 Assessment Methodology

To assess anticipation strength, we consider laboratory and field situations where the primary base state is sampled from a certain distribution and its evolution at a certain step time is observed. This distribution induces a distribution of $d\mu_q$ defined on a probability space of spectral measures. In our analyses we focus on distributions conditional under $dx_{q,0} + dx_{q,1} = dv_q$, from which any other distributions are obtained as products with the distribution of $dv_q$. In this conditional model only the spectral difference measure $dx_s$ is stochastic, whereas $dv_q$ is pre-scribed.

In contrast to (Thomann, 2008), where we asserted, that technologies for the preparation of orthogonal evolutions of a given step size were likely to exhibit orthogonal anticipation of significant strength, we do not make here any specific assumptions on the distribution of the spectral difference, but will make reference to its mean $y_1$ and variance $y_2 = E((y - y_1)^2)$ under the given distribution, which by the boundedness of $y$ always exist.

The following definitions are used below:

Definition 5:

a) $P_N = \sum_{0 \le n < N} p_n$ is the probability to measure one of the states $q_{nT}$ $(0 \le n \le N)$ at time $\frac{T}{2}$,
b) $\langle |n|^r \rangle_N \stackrel{\text{def}}{=} \sum_{0 \le n < N} |n|^r p_n$. The special case $r = 1$ is the *anticipation look-ahead*.

■

### 2.4 Anticipation Strength

In this chapter we determine the anticipation strength for model and general type states and measures with $L \le \infty$. We focus on positive evolutions with well-behaved measures and their embedded orthogonal evolutions. These results generalize those of (Thomann, 2008).

Theorem 4: Anticipation Strength

Define $\kappa_{n,m} = \frac{n+m/N}{L}$ $(0 \le n < L, 0 \le m \le M, 0 < M \le N)$.

a) Let $\mu_q$ be well-behaved and $q_0$ have positive evolution of order $L$ and size $\zeta$. Then
   1. $P_L \le \zeta^2 \|y\|_{2,dv_s}^2 \le \zeta^2 \le 1$
   2. $E(P_L) \le \zeta^2 \left(\|y_1\|_{2,dv_s}^2 + L\|y_2\|_{1,dv_{s,pp}^2}\right)$, where $dv_{s,pp}$ is the pure point component of $dv_s$ and $dv_{s,pp}^2$ its square.
b) Let $y$ be such that $x_s\left(\left[\kappa_{n,m}, \kappa_{n,m+1}\right)\right) = \frac{c}{LM}$ for some $0 < L$, $0 < c \le 1$, $0 < M = \varepsilon N$ and $0 < \varepsilon < 1$. Then for $L \to \infty$ and $N \to \infty$

1. $p_n \approx \left(\frac{c}{\sin \pi\left(n-\frac{1}{2}\right)/L} \frac{\sin \varepsilon\pi\left(n-\frac{1}{2}\right)/L}{\varepsilon\pi\left(n-\frac{1}{2}\right)}\right)^2$ with maximum value $\approx \frac{4c^2}{\pi^2}$ at $n \in \{0,1,L-1,L\}$, minimum value $\approx 4\varepsilon^{-2}\pi^{-2}L^{-2}c^2 \sin^2\frac{\varepsilon\pi}{2} \to L^{-2}c^2\ (\varepsilon \to 0)$ at $n = \left[\frac{L}{2}\right]$, and $p_n \in O(n^{-2})\ \left(n \to \left[\frac{L}{2}\right]\right)$.
2. $\langle n^1 \rangle = \ln L + O(1)$, $\langle n^r \rangle = O(L^{r-1})\ (r > 1)$.
3. $P_L = c^2\left(1 + O(L^{-1})\right)$.

c) Let $y$ be such that $x_s([\kappa_{nm}, \kappa_{n,m+1}) = \frac{c(-1)^n}{LM}$ for some $0 < L$, $0 < c \leq 1$, $0 < M = \varepsilon N$ and $0 < \varepsilon < 1$. Then for $L \to \infty$ and $N \to \infty$

1. $p_n \approx \left(\frac{c}{\cos \pi\left(n-\frac{1}{2}\right)/L} \frac{\sin \varepsilon\pi\left(n-\frac{1}{2}\right)/L}{\varepsilon\pi\left(n-\frac{1}{2}\right)}\right)^2$ with minimum value $\approx L^{-2}c^2$ at $n \in \{0,1,L-1,L\}$ , maximum value $\approx 16\varepsilon^{-2}\pi^{-4}c^2 \sin^2\frac{\varepsilon\pi}{2} \to \frac{4c^2}{\pi^2}\ (\varepsilon \to 0)$ at $n = \left[\frac{L}{2}\right]$, and $p_n \in O\left(\left(\frac{L}{2} - n\right)^{-2}\right)\ \left(n \to \frac{L}{2}\right)$.
2. $\langle n^r \rangle = O(L^r)\ (r > 1)$.
3. $P_L = 4\varepsilon^{-2}\pi^{-2}c^2 \sin^2\frac{\varepsilon\pi}{2}\left(1 + O(L^{-1})\right) \to c^2\ (\varepsilon \to 0, L \to \infty)$.

Proof:

Part a.1) By theorem 2, $s_0$ is the embedded orthogonal evolution. As $s_0, \ldots, s_L$ are mutually orthogonal, $P_L = \|z\|^2_{2,dv_s}$ , where $z$ equals the projection of $y$ on $\mathrm{Span}\{s_0, \ldots, s_L\}$.

Part a.2) By (Last, 1996) $v_s$ can be decomposed in a continuous and a pure point part, which are mutually singular. Therefore the second integral in the decomposition $E(P_L) = \iint_{\kappa \neq \kappa'} e^{-i\left(n+\frac{1}{2}\right)(\kappa-\kappa')} y_1(\kappa) y_1(\kappa') dv_s(\kappa) dv_s(\kappa') + \int E\left(y^2(\kappa)\right) dv_s(\kappa) dv_s(\kappa)$ only depends on $dv^2_{s,pp}$. The result follows from the identity $y_2 = E(y^2) - y_1^2$. Notice that, if the evolution is of order $L$, then $dv_{s,pp}$ is either zero or at least of dimension $L+1$, thus $(L+1)\int dv^2_{s,pp} \leq \left(\int dv_{s,pp}\right)^2$.

Part b.1) As $N \to \infty$ the anticipation amplitudes approximate
$\alpha_n = \sum_{m=0}^{L-1} \int_{-\varepsilon/2}^{\varepsilon/2} c e^{-2\pi i\left(n-\frac{1}{2}\right)(m+\kappa)/L} \frac{d\kappa}{2\pi} = c \int_{-\varepsilon/2}^{\varepsilon/2} e^{-2\pi i\left(n-\frac{1}{2}\right)\kappa/L} \frac{d\kappa}{2\pi} \sum_{m=0}^{L-1} e^{-2\pi i\left(n-\frac{1}{2}\right)m/L}$.
The maxima, minima and asymptotics are elementary.

Part b.2) From b.1) by the Euler summation formula.

Part b.3) By Lemma 2 below.

Part c.1) In the same way as part b.1).

Part c.2) From c.1) and c.3), as $\langle n^r \rangle \geq \left(\frac{L}{4}\right)^r P_L$.

Part c.3) By Lemma 1 below.

■

Notice in parts b-c that $x_s$ approximates an evolution of order $L$, but this $L$ is unrelated to the order of the evolution under $\mu_q$. The following lemma is obtained by direct summation, adapting theorem 4.9a of (P.Henrici, 1988) to periodic functions.

<u>Lemma 2</u>

$$\sum_{n=0}^{p-1} p^{-2} \sin^{-2} \pi\left(n - \tfrac{1}{2}\right)/p = 1.$$

## 3 Conclusions

We have classified and analyzed quantum evolutions with arbitrary spectrum and shown in theorem 1 that every positive evolution (see definition 2.c) possesses an embedded orthogonal evolution. Our findings on eigenvalues, spectra and domains of positivity have been summarized in theorem 2 for the finite-dimensional and in theorem 3 for the infinit-dimensional case. Anticipation and retrospection have been introduced (see definition 3) for positive evolutions, and their strength has been assessed in theorem 4 in the general setting of well-behaved measures (see definition 4.c).

Our analyzes show that anticipation and retrospection occur in a wide class of quantum evolutions, generalizing and extending (Thomann, 2008) where only equally distributed measures were considered.

An interesting open question is how systems exhibiting anticipation and retrospection interact with other systems, particularly with the environment. In the presence of einselection (Zurek, 2003), interesting phenomena may occur. E.g., if decoherence sets in randomly in the interval $[0, LT]$, then theorem 4.a.1. implies $\langle |n| \rangle_L \leq \frac{L-1}{2} \zeta^2 \|y\|_{2,dv_s}^2$. Given suitable timing, a Zeno-like effect may occur, but due to anticipation and retrospection with rapid evolution forward or even backward in time.

While the domain of positivity has been analyzed in our study, the requirement of positive evolution in the definition of anticipation and retrospection raises the demand for the demonstration of the presence of positive evolution in physical model systems. This will be the subject of a future paper, presenting numerical simulations showing that anticipation and retrospection occur in a variety of spectra including that of the H-atom. The quantum anticipation simulator software running und Microsoft Windows is already downloadable from the author's homepage
http://www.thomannconsulting.ch/public/aboutus/aboutus-en.htm.

## Appendix A: Spectral-analytic Lemmata

This appendix contains some useful lemmas from spectral analysis which are referenced in the main body of this paper. Our proofs are original, and we do not know of any references in the literature for them.

### A.1 Inversion of the Discrete Fourier Transform $\mathbf{d\nu}^{\wedge}(\mathbf{n})$ $(\mathbf{n} \in \mathbb{Z})$

Integrating any finite measure $\mu$ yields a continuous function $f$. As the reduced measure $d\nu$ has support in the compact set $K$, it depends only on $F = f|K$, which in turn can be represented by a Fourier series. Explicit formulae for the coefficients of this series in terms of the discrete Fourier transform $d\nu^{\wedge}(n)$ are given below.

Lemma A.1: Inversion of the Discrete Fourier Transform $\mathbf{d\nu}^{\wedge}(\mathbf{n})$ $(\mathbf{n} \in \mathbb{Z})$

Let $\nu$ be a finite complex measure supported in the half-open interval $K = [-\pi,\pi)$, and $\beta_n = d\nu^{\wedge}(n)$ $(n \in \mathbb{Z})$ be given, where $d\nu^{\wedge}(t) = \int e^{-i\kappa t}\, d\nu(\lambda)$. Then

a) There is an absolutely continuous function $F$ supported on $K$ such that
$$d\nu = D^2 F - \beta_0 \delta_\pi - \gamma_0 D \delta_\pi,$$
where the derivative $D = 2\pi i \frac{\partial}{\partial\kappa}$ is taken in the distributional sense.

b) $(2\pi)^2 F = B(\kappa) + c - 2\pi i \gamma_0 \kappa - \beta_0 \frac{\kappa^2}{2}$ $(|\kappa| \leq \pi)$, where $B(\kappa) = \sum_{n\neq 0} \frac{\beta_n}{n^2} e^{-in\kappa}$, $\gamma_0 = \frac{\partial}{\partial\kappa} B(-\pi) + \beta_0 \pi$ and $c = -B(-\pi) - i\gamma_0 \pi + \beta_0 \frac{\pi}{4}$.

c) For any test function $\phi$
$$2\pi i \nu \cdot \phi = \sum_{n\neq 0} \frac{\beta_n}{n} \left(e^{-in\kappa} \cdot \phi\right) + (i\beta_0 \kappa - 2\pi\gamma_0) \cdot \phi$$
$$d\nu \cdot \phi = \sum \beta_n \left(e^{-in\kappa} \cdot \phi\right)$$

Proof:

All references qualified by an "R" in the proof below address (Rudin, 1973). Notice that, other than (Rudin, 1973), we are scaling $D$ by the factor $2\pi$, to let $K$ have unit Lebesgue measure.

W.l.o.g. $|\beta_n| \leq 1$. Define $\nu(\kappa) = \nu\big([-\infty,\kappa)\big)$. Then $\nu(-\pi) = 0$ and $\nu(\pi) = \beta_0$. Let $\gamma_n = \int_{-\pi}^{\pi} \kappa^n \nu(\kappa) \frac{d\kappa}{2\pi}$.

First notice that $\Lambda\phi = \int \phi d\nu$, defined for test functions $\phi$, is a functional of zero order in the sense of the note following theorem R6.8, as $|\Lambda\phi| \leq \max_{\kappa\in K} |\phi|$. By theorem R6.27 and exercise R6.17 there is a continuous function $f \in C(\mathbb{R})$ such that $d\nu = D^2 f$ in the distributional sense, i.e. $D\Lambda\phi = -\Lambda D\phi$ (see R6.12). To adjust for the factor $2\pi i$, $Df = -i(2\pi)^{-1}\nu$. W.l.o.g. $f(\kappa) = 0$ $(\kappa \leq -\pi)$.

We represent $F = f|K$ as a function on $\mathbb{R}$ by defining $F = f(1 - H_\pi)$, using the Heaviside function $H_\pi = [\kappa \geq \pi]$. With $DH_x = 2\pi i\delta_x$, $g\delta_x = g(x)\delta_x$ and $f(\pi) = \int_{-\pi}^{\pi} -i\,\nu(\kappa)\frac{d\kappa}{(2\pi)^2} = -i(2\pi)^{-1}\gamma_0$ we find

(1.1) $$F = f(1 - H_\pi)$$
(1.2) $$DF = -i(2\pi)^{-1}\nu(1 - H_\pi) - 2\pi i f\delta_\pi = -i(2\pi)^{-1}\nu(1 - H_\pi) - \gamma_0\delta_\pi$$
(1.3) $$D^2F = (1 - H_\pi)d\nu - \nu\delta_\pi - \gamma_0 D\delta_\pi = d\nu - \left(\beta_0 + 2\pi i\gamma_0\frac{\partial}{\partial\kappa}\right)\delta_\pi$$

The products $f(1 - H_\pi)$ and $\nu(1 - H_\pi)$ are functionals, have been differentiated using the product rule, which is not generally applicable (see R6.14 and R6.15). However, integration by parts of the r.h.s. of $Df(1 - H_\pi)\cdot\phi = -Df(1 - H_\pi)\cdot\phi' = -\int_{-\pi}^{\pi} f\phi'$ yields the r.h.s. of 1.2, and $D\left(-i(2\pi)^{-1}\nu(1 - H_\pi)\right)\cdot\phi = -\nu(1 - H_\pi)\cdot\phi' = -\int_{-\infty}^{\infty}\nu\phi' + \int_{\pi}^{\infty}\nu\phi' = (d\nu - \nu\delta_\pi)\cdot\phi.$

This proves part a).

Now by theorem 7.15, the Fourier transform $(D^2F)^\wedge(t) = (2\pi)^2t^2F^\wedge$, by example 7.16.3, example 7.16.5 and theorem 7.2a) $\delta_\pi^\wedge = (2\pi)^{-1}e^{-i\pi t}$ and $(D\delta_\pi)^\wedge = te^{-i\pi t}$, thus

(2) $$F^\wedge = \frac{d\nu^\wedge - (\beta_0 + 2\pi\gamma_0 t)e^{-i\pi t}}{(2\pi)^2t^2}.$$

As $F$ is continuous on a compact set (the closure of $K$), the Fourier series

(3) $$F = \sum b_n e^{-in\kappa}$$

exists, and its Fourier transform

(4) $$F^\wedge(t) = \sum b_n\frac{\sin\pi(t-n)}{\pi(t-n)}.$$

Combined with (6) this yields

(5) $$b_n = F^\wedge(n) = \frac{\beta_n - (-1)^n(\beta_0 + 2\pi n\gamma_0)}{(2\pi)^2n^2}.$$

The singularity at the origin is removable; $d\nu^\wedge$ is analytic, as $d\nu$ has support in $K$ and $t^{-1}d\nu^\wedge(t) = \nu^\wedge(t) = \gamma_0$ $(t = 0)$. Therefore $b_0$ is well defined.

We insert (5) into (3) and sum the terms on the r.h.s. of (5) up separately into the absolutely convergent series

(6) $$B(\kappa) = \sum_{n\neq0}\frac{\beta_n}{n^2}e^{-in\kappa}$$

and the two sums considered next. As $i\frac{\partial}{\partial\kappa}\sum_{n\neq0}\frac{(-1)^n}{n^2}e^{-in\kappa} = \sum_{n\neq0}\frac{(-1)^n}{n}e^{-in\kappa} = \ln\frac{1+e^{-i\kappa}}{1+e^{i\kappa}} = \ln e^{-i\kappa} = -i\kappa$ $(|\kappa| < \pi)$, the absolutely convergent series on the l.h.s. equals $c - \frac{\kappa^2}{2}$ for some constant $c$. Redefining $c = c + b_0$ we obtain

(7.1) $$(2\pi)^2F = B(\kappa) + c - 2\pi i\gamma_0\kappa - \beta_0\frac{\kappa^2}{2}$$
(7.2) $$2\pi\nu\cdot\phi = i\sum_{n\neq0}\frac{\beta_n}{n}\left(e^{-in\kappa}\cdot\phi\right) + (2\pi i\gamma_0 + \beta_0\kappa)\cdot\phi$$

(7.3) $$dv \cdot \phi = \sum_{n \neq 0} \beta_n \left( e^{-in\kappa} \cdot \phi \right) + \beta_0 \cdot \phi$$

(7.4) $$2\pi i \gamma_0 = \frac{\partial}{\partial \kappa} B(-\pi) + \beta_0 \pi$$

(7.5) $$c = -B(-\pi) - 2\pi^2 i \gamma_0 + \beta_0 \frac{\pi^2}{2}$$

where (7.1) is defined for $|\kappa| \leq \pi$ due to the continuity of $F$, (7.2) and (7.3) for test functions $\phi$ with support in $K$.

(7.1) follows from the norm convergence of $B$, (7.2) and (7.3) from 6.17, as $B \cdot \phi = \sum_{n \neq 0} \frac{\beta_n}{n^2} \left( e^{-in\kappa} \cdot \phi \right)$ exists for every $\phi$ due to the norm convergence of $B$ and boundedness of $e^{-in\kappa}$.

This proves part b-c).

(7.4) and (7.5) are determined from the boundary condition $\nu(-\pi) = f(-\pi) = 0$. $F$ is absolutely continuous if $\nu$ is continuous, but may not even be one-sided differentiable in the singular-continuous spectrum and at points of accumulation of the discontinuous spectrum. In that case the r.h.s. of (7.4) cannot be evaluated. $\gamma_0$ may be approximated from (7.2) using a sequence of test functions converging to $H_{-\pi} - H_{\pi}$. $d\nu$ is always well defined by (7.3).

■

The following examples demonstrate the power of lemma 1.

Examples

1) Infinite orthogonal evolution $\beta_n = \delta_n$.
   $B = 0$, $F = \frac{1}{8}\left(-1 - \frac{2\kappa}{\pi} - \frac{\kappa^2}{\pi^2}\right)$, $\nu = \frac{1}{2}\left(1 + \frac{\kappa}{\pi}\right)$, $d\nu = 1$.
2) Periodic evolution $\beta_n = \delta_{n \bmod p}$.
   For $0 \leq \lambda = \kappa + \pi - 2\pi m p^{-1} < 2\pi p^{-1},\ 0 \leq m < p$:
   $\frac{\partial}{\partial \kappa} B = \sum_{n \neq 0} \frac{-i}{np} e^{-inp\kappa} = ip^{-1} \ln \frac{1 - e^{-ip\kappa}}{1 - e^{+ip\kappa}} = ip^{-1} \ln e^{-i(\pi + p\kappa)} = ip^{-1} \ln e^{-i(\pi + p\lambda)} = \lambda + p^{-1}\pi$. Inserting this into (11.2) yields $\nu = mp^{-1}$, as the $\lambda$ term is cancelled out and the boundary condition at $-\pi$ is met. (11.3) yields $B = \sum_{n \neq 0} e^{-inp\kappa}$ which for $p\kappa \neq 0 \bmod 2\pi$ equals $\frac{e^{-ip\kappa}}{1 - e^{-ip\kappa}} + \frac{e^{ip\kappa}}{1 - e^{ip\kappa}} = -1$, i.e. $d\nu = B + 1 = 0$, but at the singularities equals (by lemma A.2.c below) the $\delta$ functional, so $d\nu = \sum_{0 \leq m < p} p^{-1} \delta_{2\pi m p^{-1}}$.
3) Finite or infinite point spectrum $\beta_n = \sum \alpha_m e^{in\kappa_m}$, $\sum |\alpha_m| = 1$.
   By (7.3) and example 2, $d\nu = \beta_0 + \sum_{n \neq 0} \sum_m \alpha_m e^{-in(\kappa - \kappa_m)} = \beta_0 + \sum_m \alpha_m \left( \delta_{\kappa_m} - 1 \right) = \sum_m \alpha_m \delta_{\kappa_m}$.

## A.2 Representation of the Dirac δ-Functional

This lemma establishes a representation of the Dirac $\delta$-functional for a wide range of measures, including Lebesgue and pure point measure. It complements theorem 6.32 from (Rudin, 1973).

Recall from (Last, 1996) that a positive measure $\mu$ is $\alpha$-Hölder-continuous at $x$, if $\limsup_{\delta\to 0}\frac{\mu((x-\delta/2,x+\delta/2))}{\delta^\alpha}<\infty$ for some $0\le\alpha\le 1$, $\alpha$-Hölder-continuous in a set $A$ if $\alpha$-Hölder-continuous for all $x\in A$ and uniformly $\alpha$-Hölder-continuous ($U\alpha H$) in $A$, if $\limsup_{\delta\to 0}\frac{\mu((x-\delta/2,x+\delta/2))}{\delta^\alpha}<C<\infty$ for all $x\in A$.

Lemma A.2: Representation of the Dirac δ-Functional

Let $\mu$ be a finite, positive measure $\alpha$-Hölder-continuous at the origin and define the functional $|\delta|$ by $|\delta|\,\phi=|\phi(0)|$ for any test function $\phi$. For any test function $\Psi$ supported in $I=\left(-\frac{1}{2},+\frac{1}{2}\right)$ define $\Psi_N(\kappa)=\Psi(N\kappa)$. Then for $0\le c=\limsup_{N\to\infty}\left|\int N^\alpha\Psi_N(\kappa)d\mu(\kappa)\right|<\infty$

(a) $\limsup_{N\to\infty}|N^\alpha\Psi_N d\mu|=c|\delta|$, in the sense that, for any test function $\phi$: $\limsup_{N\to\infty}\left|\int N^\alpha\Psi_N(\kappa)\phi(\kappa)d\mu(\kappa)\right|=c|\phi(0)|$.

(b) If $\alpha=1$ and $d\mu(\kappa)=f d\kappa$, where $f$ is continuous at the origin, then the limit $\lim_{N\to\infty}N\Psi_N d\mu=\mathrm{c}\delta$ exists and $c=f(0)\int\Psi\,d\kappa$.

(c) If $\alpha=0$, then $\lim_{N\to\infty}\Psi_N d\mu=c\delta$ exists and $c=\Psi(0)\mu(\{0\})$.

Proof:

Part a)

As $\Psi_N$ has support $I_N=\left(-\frac{1}{2N},+\frac{1}{2N}\right)$, the integration domain is $I_N$. Therefore $\limsup_{N\to\infty}\left|N^\alpha\int\Psi_N(\kappa)\phi(\kappa)d\mu(\kappa)\right|=\limsup_{N\to\infty}\left|\phi(0)N^\alpha\int\Psi_N(\kappa)d\mu(\kappa)\right|=c|\phi(0)|$, where $c=\limsup_{N\to\infty}\left|\int N^\alpha\Psi_N(\kappa)d\mu(\kappa)\right|\le\limsup_{N\to\infty}N^\alpha\int|\Psi_N(\kappa)|d\mu(\kappa)\le\|\Psi\|_{\infty,d\kappa}\limsup_{N\to\infty}N^\alpha\mu(I)<\infty$.
The last inequality follows from the premises.

Part b) holds as by definition $d\mu(\kappa)=f(\kappa)d\kappa\approx f(0)d\kappa$ ($\kappa\in I_N$).

Part c) is a consequence of the definitions and $\int\Psi_N d\mu=\int_{\kappa\neq 0}\Psi_N d\mu+\Psi(0)\mu(\{0\})$, where the first integral vanishes in the limit due to the contraction of $\mathrm{I}_N$.

■

Corollary A.2

Let $\mu$ and $I$ be as in lemma A.2, $\mu$ supported in $I$. Then

(a) There is a constant $0\le c\le\infty$ such that $\limsup_{N\to\infty}N^{\alpha-1}\left|\sum_{|n|<N/2}e^{2\pi in\kappa}\,d\mu(\kappa)\right|=c|\delta|$, in the sense that $\limsup_{N\to\infty}N^{\alpha-1}\left|\int\frac{\sin\pi N\kappa}{\sin\pi\kappa}\phi(\kappa)d\mu(\kappa)\right|=c|\phi(0)|$ for any test function $\phi$ supported in $I$.

(b) If $\alpha=1$ and $d\mu(\kappa)=f d\kappa$, where $f$ is continuous at the origin, then the limit $\lim_{N\to\infty}N^{\alpha-1}\sum_{|n|<N/2}e^{2\pi in\kappa}\,d\mu(\kappa)=\mathrm{c}\delta$ exists and $c=f(0)$.

(c) If $\alpha=0$ and the origin is an isolated point of the support of $\mu$, then $\lim_{N\to\infty}N^{\alpha-1}\sum_{|n|<N/2}e^{2\pi in\kappa}\,d\mu(\kappa)=\mathrm{c}\delta$ exists and $c=\mu(\{0\})$.

(d) If $\mu$ is supported in $\mathbb{R}$ and $\alpha$-Hölder-continuous at $n\in\mathbb{Z}$ then $\limsup_{N\to\infty}N^{\alpha-1}\left|\sum_{|n|<N/2}e^{2\pi in\kappa}\,d\mu(\kappa)\right|=\sum_{n\in\mathbb{Z}}c_n|\delta_{\kappa-n}|$.

Proof:
Part a)
If $\frac{1}{2} \geq \kappa \geq \varepsilon > 0$, then $\sin \pi\kappa \geq \sin \pi\varepsilon$ and $N^{\alpha-1} \int_{\varepsilon}^{1/2} \sin \pi N\kappa \, d\mu(\kappa) \in O(N^{\alpha-1})$ for $\alpha < 1$, and for $\alpha = 1$: $\int_{\varepsilon}^{1/2} \sin \pi N\kappa \, d\kappa = \left| -\frac{\cos \pi N\kappa}{\pi N} \in O(N^{-1}) \right.$. Thus this part of the integral vanishes.

If $0 \leq \kappa < \varepsilon$, then $\sin \pi\kappa = \pi\kappa + O(\varepsilon^2)$. Therefore $N^{\alpha-1} \frac{\sin \pi N\kappa}{\sin \pi\kappa} \approx N^{\alpha} \frac{\sin \pi N\kappa}{\pi N\kappa} \approx N^{\alpha}$. As the integral vanishes for $|\kappa| > \varepsilon$, the proof of lemma A.2.a extends to $\frac{\sin \pi\kappa}{\pi\kappa}$. However, $c$ may be infinite (see remark below).

Part (b) follows from $\int \frac{\sin \pi\kappa}{\pi\kappa} d\kappa = \frac{2}{\pi} \mathrm{Si}(\infty) = 1$.

Part (c) follows from $\frac{\sin N\pi\kappa}{N\pi\kappa} \to 1 \ (\kappa \to 0)$ and $N^{-1} \sum_{|n|<N/2} 1 = 1 + O(N^{-1})$.

Part (d) is an immediate consequence of (a) and the periodicity of $\frac{\sin \pi N\kappa}{\sin \pi\kappa}$.

■

Even uniform $\alpha$-Hölder-continuity of $\mu$ in a neighborhood of the origin is not sufficient for finiteness of the upper bound in corollary A.2.a above, as it does not imply more than $\left| \int_{-\varepsilon}^{\varepsilon} N^{\alpha} \frac{\sin \pi N\kappa}{N\pi\kappa} d\mu(\kappa) \right| \leq \sum_{|n|<N\varepsilon} \frac{1}{n} \cdot \frac{\mu\left(-\frac{N^{-1}}{2}, +\frac{N^{-1}}{2}\right)}{N^{-\alpha}} \in O(\ln N)$. On the other hand, the Cantor measure and many other self-similar measures, as well as the conditions of corollary A.2.b-c, yield finite bounds. Thus this corollary is widely applicable.

### A.3 Time-averages of $|d\mu\hat{}(t)|^2$

(Last, 1996) analyzes the time average $\langle |d\mu\hat{}(t)|^2 \rangle_{T^{1-\alpha}} = T^{\alpha-1} \int_{-T/2}^{T/2} |d\mu\hat{}(t)|^2 dt$ and proves in theorem 2.5 the following result first established by (Strichartz, 1990):

Let $\mu$ be a finite $U\alpha H$ measure and $f \in \mathcal{L}^2(d\mu)$. Then there is a constant $C$ only depending on $\mu$ such that $\langle |(f d\mu)\hat{}(t)|^2 \rangle_{T^{1-\alpha}} < C \|f\|_{2,d\mu}^2$.

Notice however, that these are asymptotic results. In evolutions of order $L$ theorem 4.a.1 implies $\int_{-T/2}^{T/2} |d\mu\hat{}(t)|^2 dt \leq 1$ for $T = L$, while the l.h.s. diverges for $\alpha < 1$ and $T \to \infty$ by Strichartz' theorem.